\documentclass[aps,prl,twocolumn,showpacs]{revtex4}
\usepackage{graphicx}
\usepackage{float}

\begin{document}



\title{Fermi Surface Reconstruction in CeTe$_2$ Induced by
	Charge Density Wave:\\
	 ARPES Study}

\author {J.-S. Kang$^1$, D. H. Kim$^1$, H. J. Lee$^1$, J. H. Hwang$^1$,
	Han-Koo Lee$^2$, H.-D. Kim$^2$, \\
	B. H. Min$^3$, K. E. Lee$^3$, Y. S. Kwon$^3$, 
	J. W. Kim$^4$, Kyoo Kim$^4$, B. H. Kim$^4$, and B. I. Min$^4$}

\affiliation{$^1$Department of Physics, 
	The Catholic University of Korea, Bucheon 420-743, Korea}

\affiliation{$^2$Pohang Accelerator Laboratory, POSTECH, 
	Pohang 790-784, Korea}

\affiliation{$^3$Department of Physics, Sungkyunkwan University,
        Suwon 440-746, Korea}

\affiliation{$^4$Department of Physics, Pohang University of Science 
	and Technology, Pohang 790-784, Korea}

\date{\today}

\begin{abstract}

Electronic structures of a charge-density-wave (CDW) system 
CeTe$_{2-x}$Sb$_x$ ($x$=0, 0.05) have been investigated 
by employing angle-resolved photoemission spectroscopy (ARPES). 
The observed Fermi surface (FS) agrees very well with 
the calculated FS for the undistorted CeTe$_2$ both in shapes 
and sizes. The metallic states crossing the Fermi level 
($\rm E_F$) are observed in ARPES. The carriers near $\rm E_F$ 
have mainly the Te(1) $5p$ character, with the negligible 
contribution from Ce $4f$ states to the CDW formation. 
The supercell (shadow) bands and the corresponding very weak 
FS's are found to arise from band-folding due to the interaction 
with Ce-Te(2) layers.
This work shows that the origin of the CDW formation in CeTe$_2$
is the FS nesting and that the CDW modulation vector is 
along $\Gamma$-X (${\bf Q}_{CDW}$$\approx$X).

\end{abstract} 



\maketitle

\begin{figure}[b]
\includegraphics[angle=0,width=7cm]{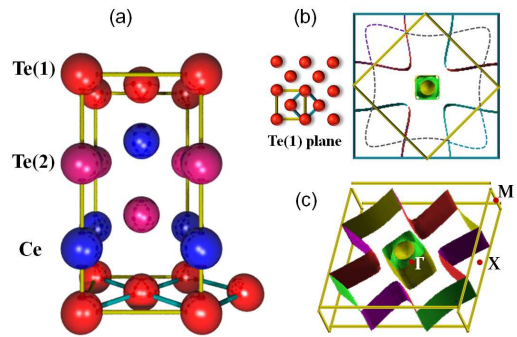}
\caption{(Color online) 
	(a) Tetragonal crystal structure of CeTe$_2$ in the non-CDW 
	phase. The unit cell of Te(1) square net is doubled 
	in tetragonal CeTe$_2$. 
	(b) The Brillouin Zone (BZ) of Te(1) sheets (outer square)
	is reduced to half (inner square) because of the larger cell 
	size of Ce-Te(2) layers.
	(c) Fermi surface of CeTe$_2$ in the tetragonal BZ of 
	the non-CDW phase, corresponding to the inner square 
	in (b).\cite{Jwkim11} 
	$\Gamma$, X, and M represent {\bf k}=(0, 0, 0),
	$\frac{2\pi}{a}$(1/2, 0, 0), and $\frac{2\pi}{a}$(1/2, 1/2, 0),
	respectively.  }
\label{cst}
\end{figure}

The charge-density-wave (CDW) formation is one of the most interesting 
phenomena in solid state physics.\cite{Voit00,Kasuya00,Yokoya01} 
CeTe$_2$ is known as a CDW system having a high CDW transition 
temperature of T$_{CDW}$$\sim$1000 K, and the CDW state in CeTe$_2$ 
coexists with magnetism and also with superconductivity 
(T$_C$=2.7 K) under pressure.\cite{Jungprb03} 
CeTe$_2$ crystallizes in the quasi-two dimensional layered 
Cu$_2$Sb-type tetragonal structure with two types of Te sites: 
Te(1) and Te(2). Te(1) atoms form planar square sheets, which 
are sandwiched along the $c$ axis by the corrugated double 
layers of Ce and Te(2) atoms (Fig.~\ref{cst}). 
The ionic configuration of CeTe$_2$ is considered to be 
Ce$^{3+}$Te(2)$^{2-}$Te(1)$^{1-}$, so that hole carriers are 
produced in Te(1) sheets.\cite{Kang06} Then the square net 
of Te(1) would be easily distorted by the Peierls-like 
mechanism\cite{Burdett83} due to the partial filling. 
Underneath this picture is the assumption of trivalent Ce$^{3+}$ 
states.\cite{Kang04}
Band-structure calculations indicate that the CDW instability occurs
due to the nesting between the Fermi surfaces in Te(1) square 
sheets in the $ab$ plane,\cite{Dimasi96,Kikuchi98,Shim04}
which was supported experimentally.\cite{Stowe00b} 

Due to the difficulty in growing high-quality single crystals 
for the angle-resolved photoemission spectroscopy (ARPES)
study, there have been only a few ARPES studies of CeTe$_2$.
A couple of works reported the Fermi surface (FS) topology 
in the CDW state of CeTe$_2$, as well as that in a similar CDW 
system LaTe$_2$, 
by using ARPES.\cite{Shin05,Ito07,Garcia07} 
Shin {\it et. al.}\cite{Shin05} reported that the FS topology 
of CeTe$_2$ in the $k_x$-$k_y$ plane is different from that 
of LaTe$_2$.\cite{Garcia07} The small square FS centered around 
$\Gamma$, predicted by band calculations, was not observed 
in ARPES. So they conjectured that the CDW gap $E_g$ is larger 
than $\ge 600$ meV, and that the magnitude of $E_g$ varies 
around the FS. This minimum value of $E_g \approx 600$ meV 
is much larger than $E_g \approx 100$ meV, found in another 
ARPES study.\cite{Kang06} The CDW distortion probed by 
TEM\cite{Shin05} was somewhat different from that 
in the literature,\cite{Stowe00b} which was attributed 
to the variation of the Te deficiency produced by different 
sample growth techniques. Ito {\it et. al.}\cite{Ito07} examined 
the FS along the $k_z$ axis, and observed a systematic 
intensity modulation in the spectral weight at FS.

\begin{figure}
\includegraphics[angle=0,width=7 cm]{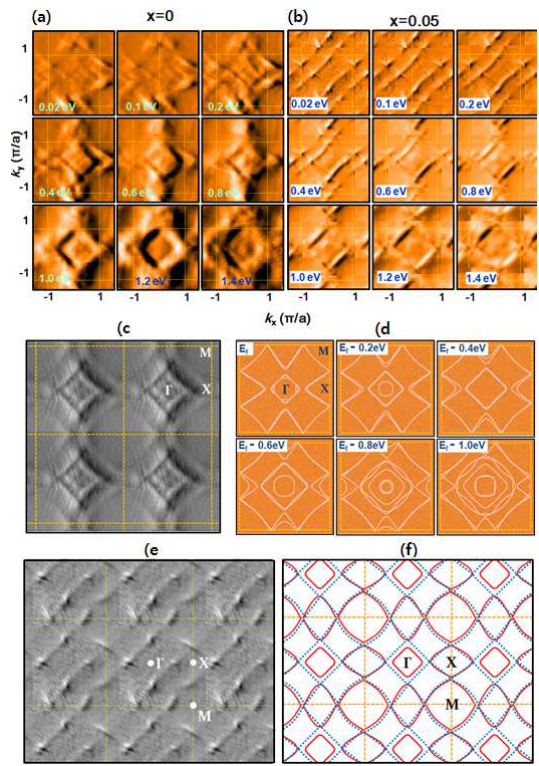} 
\caption{(Color online) 
	(a) First derivatives of the constant-energy (CE) map  
	of CeTe$_2$ with the increasing binding energy (BE)
	between BE=$0.02$ eV and BE $\sim 1.4$ eV.
	(b) Similarly for CeTe$_{1.95}$Sb$_{0.05}$.
	Here BE=$|E_i|$, where $E_i$ denotes the initial-state energy.
	Dotted lines denote the BZ.
	(c) The experimental CE map of CeTe$_2$ for BE=1.0 eV, 
	drawn in the periodic zone scheme. This map was generated 
	by the reflections of the map data that were obtained 
	for a smaller region of the BZ.
	(d) The calculated CE map of CeTe$_2$ as a function of $E_i$. 
	(e) The experimental FS map of CeTe$_{1.95}$Sb$_{0.05}$.
	(f) Comparison of the experimental FS map (blue dotted lines) 
	and calculated FS map (red solid lines).}
\label{map}
\end{figure}

The natural question is whether the FS topology and the CDW state 
of CeTe$_2$ are similar to those of LaTe$_2$ or not. 
This includes the following issues: 
(i) do the FS and the CDW distorted structure in CeTe$_2$ have 
the same symmetry as those in LaTe$_2$?, 
(ii) how large is the effect of the band folding, 
due to the interaction between Te(1) and Ce-Te(2) layers,
on the FS of Te(1) sheets?, 
and (iii) what is the CDW modulation vector, ${\bf Q}_{CDW}$?
As shown in Fig.~\ref{cst}(b), the Brillouin Zone (BZ) of Te(1) sheets 
is reduced to half, and thereby, the bands are folded into 
the reduced BZ to yield the supercell (shadow) bands. 
The FS's, denoted with dotted lines in Fig.~\ref{cst}(b), 
come from those shadow bands.
We have resolved these questions by performing careful ARPES 
measurements for high-quality stoichiometric single crystals of 
CeTe$_{2-x}$Sb$_x$ ($x=0, 0.05$).
We also report the detailed study of the FS topology, 
the energy-dependent behavior of the constant energy (CE) map, 
and the clearly dispersive feature of the states near the Fermi level
($\rm E_F$) in CeTe$_2$. 


High-quality CeTe$_{2-x}$Sb$_x$ single crystals having 
very low residual resistivity were grown by using the self-fluxed 
Bridgeman method.\cite{BHM02}
The quality and the orientation of the single crystal were checked
by Laue patterns. ARPES experiments were carried out at the 3A1 
beamline of the Pohang Light Source (PLS) with the beam size of 
$< 50~\mu$m and using a Scienta SES-2002 electron energy analyzer. 
Single crystals were cleaved {\it in situ} at $T \sim 30$ K 
under the pressure better than $5\times 10^{-11}$ Torr, 
which exposed the (001) surfaces. The Fermi level and the overall 
instrumental resolution of the system were determined from 
the Fermi edge of an evaporated Cu metal. The energy resolution 
($\Delta E$) and the momentum resolution ($\Delta k$) were 
set to be $\Delta E \sim 80$ meV and $\Delta k \approx 0.01 
\AA^{-1}$, respectively, at $h\nu \sim 110$ eV. 


Figure~\ref{map} shows the first derivatives of the CE maps
of CeTe$_2$ (Left) and CeTe$_{1.95}$Sb$_{0.05}$ (Right) 
vs. the initial-state energy ($E_i$) of 
$-1.4$ eV $\le E_i \le -0.02$ eV.
These data were obtained at $T$$\sim$ 30 K with $h\nu\approx$115 eV.
In plotting each CE map, the spectral intensity of 
$E_i \pm 100$ meV was integrated. This figure shows that 
the CE maps of $x$=0 and $x$=0.05 are essentially the same
in their shapes, sizes, and energy-dependent behavior. 
But the CE maps of $x$=0.05 are sharper than those of $x$=0. 
Thus the energy-dependent evolution of the CE map is manifested 
more clearly in $x$=0.05 than in $x$=0.\cite{05} 

The four-fold symmetry is observed in the FS\cite{FS} and CE maps 
for both $x$=0 and 0.05. 
The feature of the four-fold symmetry is shown more clearly in 
the CE map for $E_i$=$-1.0$ eV (Fig.~\ref{map}(c)),
which was obtained from a different cleave.
Two diamond-shaped contours are observed in the FS map, 
which is similar to the case of LaTe$_2$.\cite{Garcia07} 
This finding implies that Ce $4f$ electrons 
hardly contribute to the states near $\rm E_F$.
Note that the inner-diamond FS is clearly seen in both $x$=0 
and 0.05, in contrast to the case of Ref.\cite{Shin05}.
With increasing $|E_i|$, the size of the inner diamond increases, 
while that of the outer diamond remains nearly the same.
Such an energy-dependent behavior is consistent with that of 
the calculated FS's, shown in Fig.~\ref{map}(d).

Figure~\ref{map}(e) shows the FS of CeTe$_{1.95}$Sb$_{0.05}$, 
drawn in the periodic zone scheme. 
This FS map was obtained by integrating the ARPES spectra 
for $\rm E_F ~\pm$ 50 meV ($\rm E_F$$\equiv$0 eV).
The existence of the FS implies that there remain metallic states 
even below the CDW transition, producing the remnant ungapped FS.
This conclusion is supported by the calculated FS, 
shown in Fig.~\ref{map}(f).  In Fig.~\ref{map}(f), 
the experimental FS is compared with the calculated FS.
The former is obtained by interpolating the partially 
ungapped FS's and the latter is calculated for 
the non-CDW phase.\cite{Shim04}
The experimental and calculated FS's are very similar to each other
in both sizes and shapes, supporting that the measured FS 
reflects the FS of the non-CDW phase of CeTe$_2$. 
Two important features are noted here. First, this comparison 
confirms the FS nesting mechanism for the CDW formation in CeTe$_2$.
Secondly, the opening of the CDW gaps occurs only partially 
in some part of the FS, which is consistent with the semimetallic 
nature of  CeTe$_2$.\cite{Kang06,Lavagnini07,Lee08}

\begin{figure}[b]
\includegraphics[angle=270,width=6cm]{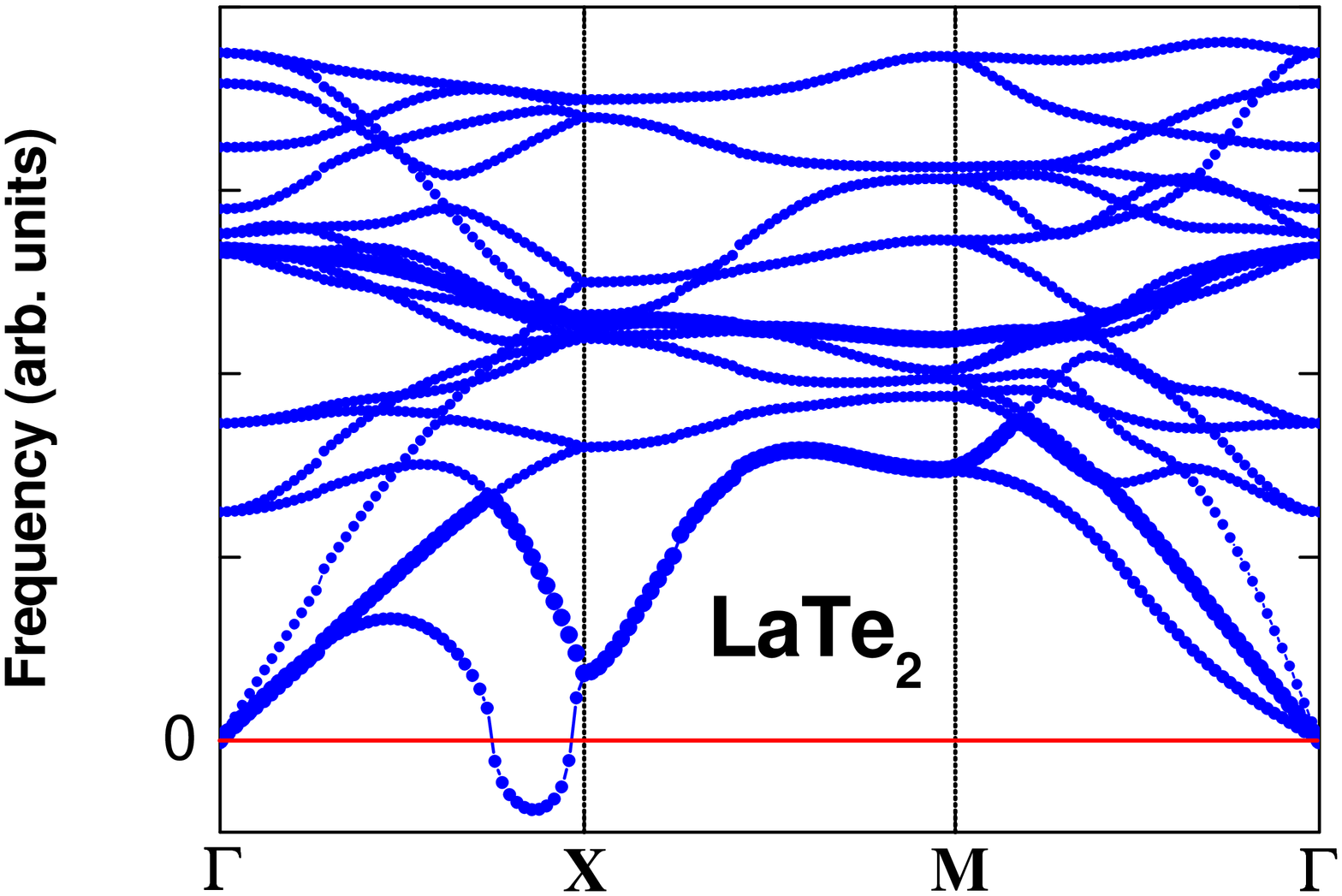} 
\caption{(Color online) 
	Phonon dispersion $\omega({\bf q})$ of undistorted LaTe$_2$. 
	Negative $\omega({\bf q})$ here represents the imaginary part 
	of the phonon frequency. Phonon softening occurs at 
	${\bf Q}_{1}$$\approx$X ($\omega^2({\bf Q}_{1}) < 0$).  }
\label{ph}
\end{figure}
 
According to band calculations, there are two FS nesting vectors,
one parallel to $\Gamma$-X $({\bf Q}_1$$\approx$X), and the other 
parallel to $\Gamma$-M $({\bf Q}_2$$\approx$M/2) for both 
LaTe$_2$\cite{Kikuchi98} and CeTe$_2$.\cite{Shim04}
In general, ${\bf Q}_{CDW}$ is determined by the FS nesting 
vector.\cite{Dimasi96} In real systems, however, the CDW 
transition occurs only when there exists a soft phonon mode 
at a specific ${\bf Q}_{CDW}$. 
In order to check the existence of a soft phonon mode in CeTe$_2$, 
we have calculated the phonon dispersion
$\omega({\bf q})$ for LaTe$_2$ for simplicity.\cite{QE}
Indeed, the calculated $\omega({\bf q})$ (Fig.~\ref{ph}) exhibits 
negative values near X. This finding manifests phonon softening 
at ${\bf Q}_1$$\approx$X, and suggests that ${\bf Q}_{CDW}$ 
in CeTe$_2$ corresponds to ${\bf Q}_1$$\approx$X, but not to 
${\bf Q}_2$$\approx$M/2.

\begin{figure}[t]
\includegraphics[angle=270,width=7.8cm]{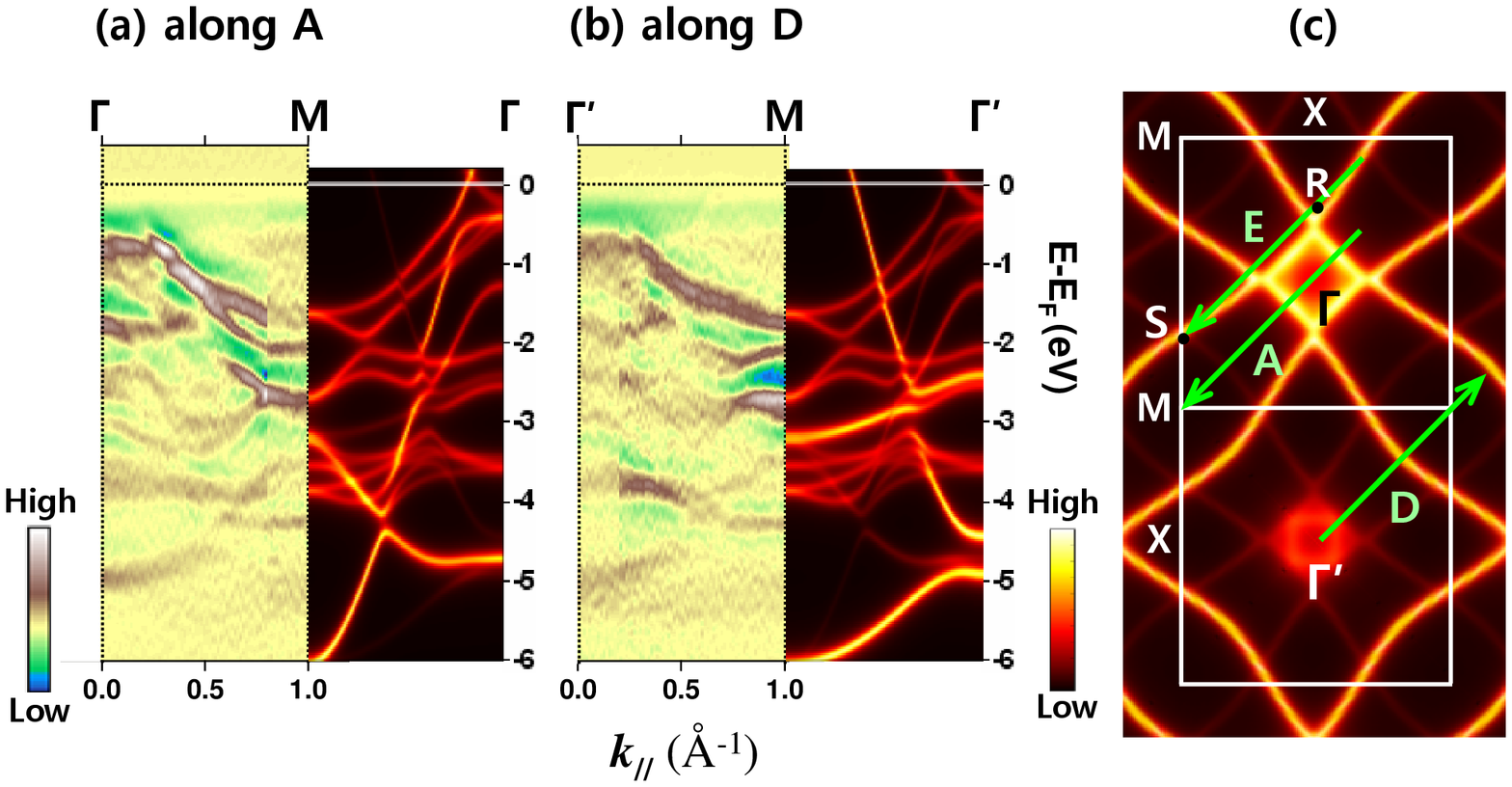}
\caption{(Color online) 
	(a) ARPES along $A$ ($\Gamma$-M), with the calculated 
	band structures along M-$\Gamma$ on the right, which is
	attached as a mirror image. 
	(b) Similarly for along $D$ ($\Gamma$-M).
	(c) Unfolded FS in the extended BZ, where two different 
	$\Gamma$-M paths, {\it i.e.} along $A$ and $D$, are indicated.
	Note that the shadow bands and the corresponding FS are 
	barely seen.  }
\label{arp}
\end{figure}

In Fig.~\ref{arp}, we have compared the ARPES intensity plots 
of CeTe$_2$ with the band structures, calculated for 
the non-CDW lattice.\cite{Jwkim11}
Left sides of Fig.~\ref{arp}(a) and (b) show the ARPES data of 
CeTe$_2$ along $\Gamma$-M in two different paths, 
along A and along D, respectively, as shown in Fig.~\ref{arp}(c).
ARPES intensity plots were made by taking second derivatives 
of the ARPES data, obtained with $h\nu\approx$104 eV.
The experimental band structures of CeTe$_{1.95}$Sb$_{0.05}$ 
(not shown here) are found to be essentially the same as those 
of CeTe$_2$ without a noticeable energy shift 
within the instrumental resolution.\cite{05}
Many dispersive bands, observed clearly in ARPES, indicate 
the good quality of the samples employed in this study. 
The overall band structures of Fig.~\ref{arp}(a) 
and (b) are similar to each other. 

But there are also some differences between two, which seem 
to arise from (i) the band-folding effect due to the increased 
unit cell (see Fig.~\ref{cst}(b)), and (ii) differences 
in the topography of the cleaved surfaces.\cite{geo}
The band-folding effect is clearly seen in the calculated band 
structures, shown on the right sides of Fig.~\ref{arp}(a) and (b).
For the calculated band structure and FS, the band unfolding 
scheme\cite{Ku10} is adopted to separate out the shadow bands 
and the corresponding FS in the larger BZ from the non-shadow 
(main) bands and the corresponding FS. In this scheme, 
the intensity of the Te(1) supercell band is proportional to 
the interaction between the Te(1) layer and the underlying 
Ce-Te(2) layer. If the interaction is weak, it results in 
the weakening of weights of supercell-folded bands and 
the corresponding FS. In Fig.~\ref{arp}(a), the wide in-plane 
Te(1) $5p$ band\cite{Shim04} starting from M at $\sim -6.2$ eV 
to near $\Gamma$ ($\sim 0.2$M) at $\rm E_F$ is prominent, 
while in Fig.~\ref{arp}(b), the in-plane Te(1) $5p$ band starting 
from $\Gamma'$ at $\sim -4.5$ eV to near M ($\sim 0.7$M) 
at $\rm E_F$ is prominent. In fact, the latter is seen as 
a dim shadow band in Fig.~\ref{arp}(a), and vice versa 
in Fig.~\ref{arp}(b).
The main and shadow bands in Fig.~\ref{arp}(a) produce 
the bright inner FS and the dim outer FS in the first BZ, 
respectively, shown in the top BZ of Fig.~\ref{arp}(c).

\begin{figure}[t]
\includegraphics[angle=270,scale=0.4]{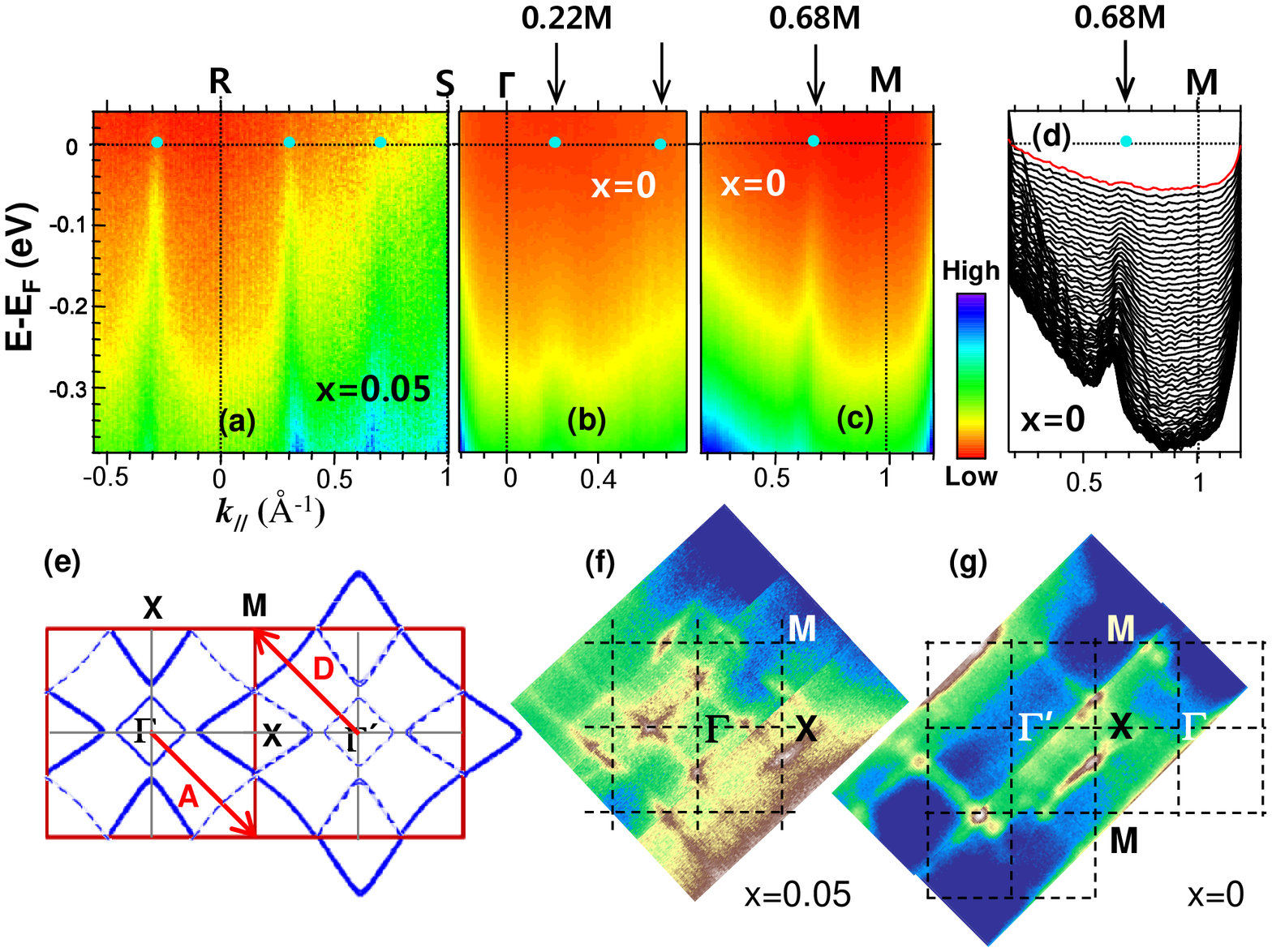} 
\caption{(Color online) 
(a) Near-$\rm E_F$ ARPES intensity plots of CeTe$_{1.95}$Sb$_{0.05}$
	along E (see Fig.~\ref{arp}(c)).
(b) Near-$\rm E_F$ ARPES of CeTe$_2$ along A ($\Gamma$-M) 
	in the first BZ.
(c) Near-$\rm E_F$ ARPES of CeTe$_2$ along D ($\Gamma^{\prime}$-M) 
	in the second BZ.
(d) Stack of MDC's, corresponding to (c). 
	The top and bottom MDC curves correspond to 
	$E_i=0$ ($\rm E_F$) and $E_i=-0.4$ eV, respectively. 
(e) Calculated FS maps for the non-CDW state of CeTe$_2$.
	Dotted lines represent the folded FS's.
(f) Experimental FS map of CeTe$_{1.95}$Sb$_{0.05}$.
(g) Similarly for CeTe$_2$.
	In (f) and (g), $\Gamma$ and $\Gamma^{\prime}$ denote 
	the $\Gamma$ points in the first BZ and the second BZ, 
	respectively.  }
\label{ef}
\end{figure}

A very good agreement is found between ARPES and calculation 
in both the dispersive feature and the energy positions of 
the bands below $-0.6$ eV. However, some discrepancies are 
also observed. Some of the calculated bands are missing in ARPES. 
In particular, the $\rm E_F$-crossing bands near $\Gamma$ and M 
in theory, originating from Te(1) $5p$-states, are not observed 
clearly in ARPES.
Only the dim features are barely seen in Fig.~\ref{arp}(a) and (b).
In order to see the bands near $\rm E_F$ more clearly, 
the ARPES intensity plots in the vicinity of $\rm E_F$ are shown 
in Fig.~\ref{ef}. 
Figures~\ref{ef}(a), (b), and (c) show the near-$\rm E_F$ 
raw ARPES spectra of CeTe$_{1.95}$Sb$_{0.05}$ along E 
and those of CeTe$_2$ along A and D in the BZ, respectively.
Figure~\ref{ef}(d) shows the stack of momentum distribution curves 
(MDC's), corresponding to the constant-energy cuts through 
the ARPES intensity plots shown in Fig.~\ref{ef}(c).

The band-crossing through $\rm E_F$ is clearly observed 
in Fig.~\ref{ef}(a). These states are expected to produce 
the FS's at the corresponding ${\bf k}_F$ values.\cite{edc}
On the other hand, in Fig.~\ref{ef}(b) and (c), there are bands 
approaching $\rm E_F$, but their spectral intensities die away 
as they approach $\rm E_F$. The expected $\rm E_F$-crossing 
positions in ARPES agree very well with those in the calculated 
bands for the non-CDW phase of CeTe$_2$ (see Fig.~\ref{arp}), 
which have mainly Te(1) $5p$ character.\cite{Shim04}  
As shown in Fig.~\ref{arp}(c), 
two bands near $\Gamma$ ($\sim 0.2M$) and M ($\sim 0.7M$) 
in Fig.~\ref{ef}(b) would produce the inner (bright) FS and 
outer (dim) FS's along $\Gamma$-M in the first BZ, respectively.
Similarly, the band near M ($\sim 0.7M$) in Fig.~\ref{ef}(c) 
would produce the larger FS along $\Gamma$-M in the second BZ. 
The much weaker spectral weight of the band near M in 
Fig.~\ref{ef}(b), as compared to that near $\Gamma$, is 
attributed to its nature of the shadow band, which originates 
from the band-folding.

The vanishing spectral intensity near $\rm E_F$ in Fig.~\ref{ef}
is considered to be related to the opening of the CDW gaps 
in some part of the FS's. This feature is revealed more clearly 
in Fig.~\ref{ef}(f) and \ref{ef}(g), which show the experimental 
FS maps of the first BZ and the second BZ, respectively, obtained 
by integrating $\rm E_F ~\pm 100$ meV. These are not the derivative 
data, but raw data. The inner FS near $\Gamma$ is certainly 
observed in Fig.~\ref{ef}(f), while the outer diamond FS is 
very weak. In contrast, the inner diamond FS near $\Gamma^{\prime}$ 
is hardly seen in Fig.~\ref{ef}(g) but the outer diamond FS is 
apparent. In view of the calculated FS's in Fig. \ref{arp}(c) 
and Fig. \ref{ef}(e), such differences can be interpreted as 
the fact that the FS's in Fig.~\ref{ef}(f) correspond to those
in the first BZ, whereas the FS in Fig.~\ref{ef}(g) corresponds
to that in the second BZ.

The feature of the vanishing spectral weight is also observed
in some part of the FS's: for example, in Fig.~\ref{ef}(f), 
the spectral weight near M is almost vanishing, 
and the outer diamond FS shows an intermittent feature.  
The polarization effect is not likely the origin of 
an intermittent feature of the outer FS since this feature is 
common for all four sides of the outer diamond. Instead, 
this region would correspond to the ${\bf k}$-points where 
CDW gaps open. Therefore the diminishing feature in 
Fig.~\ref{ef}(b) and (c) reflects the opening of the CDW gap. 
One can estimate the size of the CDW energy gap as being 
$E_g \simeq 50$ meV near M, in agreement with our previous 
finding in a different ARPES study.\cite{Kang06} 


In conclusion, the FS measured by ARPES agrees very well with 
the calculated FS for the undistorted CeTe$_2$ both in shapes 
and sizes, and the $\rm E_F$-crossing metallic states are clearly 
observed in ARPES. We have found the following answers 
to the questions addressed in the beginning:
(i) The measured FS is similar to that of LaTe$_2$,
implying that Ce $4f$ states have a minor contribution 
to the CDW formation in CeTe$_2$. 
(ii) The band-folding originating from the interaction 
with Ce-Te(2) layers produces the shadow bands and 
the corresponding FS's, which have very weak spectral weight.
(iii) The CDW modulation vector is estimated to be 
${\bf Q}_{CDW}={\bf Q}_1$$\approx$X.

This work was supported by the NRF under Contract No. 2009-0064246
and No. 2009-0079947. YSK acknowledges the NRF grant under Contract 
No. 2006-2002165 and 2009-0078025.
PLS is supported by POSTECH and MEST in Korea.


\begin{thebibliography}{99}
\bibitem{05} 
	The ground state and the electronic structures of CeTe$_2$ 
	remain almost the same in CeTe$_{1.95}$Sb$_{0.05}$ 
	in this energy scale since the Sb-concentration 
	in CeTe$_{1.95}$Sb$_{0.05}$ is very low ($2.5 \%$).
\bibitem{FS} 
	The map with $E_i$=$-0.02$ eV can be regarded as the FS
	because $\rm E_F\equiv$0 is included in this map.
\bibitem{geo} All the cleaved surfaces were mirror-shiny, but 
	had many stepped planes. Due to a finite acceptance 
	angle of the electron analyzer and a different 
	topography of a cleaved surface, each cleaved surface 
	is expected to have a different yield.
\bibitem{edc} In the energy distribution curves (EDC's), 
	the photoemission peaks that cross $\rm E_F$ were not 
	observed clearly. Nevertheless, the $\rm E_F$-crossing 
	of these bands is supported by the slope of these bands.
\end{thebibliography}
\end{document}